\documentstyle[preprint,prb,aps,epsf,eqsecnum]{revtex}
\begin{document}
\tightenlines
\preprint{ITPSB-97-66}
\draft

\title{Spin diffusion and the  spin 1/2 XXZ chain at $T=\infty$ 
from exact diagonalization}

\author{Klaus Fabricius
\footnote{e-mail Klaus.Fabricius@theorie.physik.uni-wuppertal.de}}
\address{ Physics Department, University of Wuppertal, 
42097 Wuppertal, Germany}
\author{Barry~M.~McCoy
\footnote{e-mail mccoy@insti.physics.sunysb.edu}}               
\address{ Institute for Theoretical Physics, State University of New York,
 Stony Brook,  NY 11794-3840}
\date{\today}
\maketitle

\begin{abstract}
We study the long time  behavior of
the zz and xx time dependent autocorrelation function 
of the spin 1/2 XXZ chain at $T=\infty$ 
by exact diagonalizations on a chain of 16 sites. We find that the
numerical results for the zz correlation are very well fit by 
the formula $t^{-d}[A+Be^{-\gamma (t-t_0)}\cos
\Omega(t-t_0)].$  From this we estimate $d$ as a function of the
anisotropy of the chain and study the crossover from ballistic to
diffusive behavior.

\end{abstract}
\pacs{PACS 75.10.Jm, 75.40.Gb}
\section{Introduction}

The phenomenological theory of spin diffusion was first proposed by
Bloembergen~\cite{rb}~ and De Gennes~\cite{rdg,rdgb} to give a 
description of inelastic neutron scattering in
magnetic systems at elevated temperatures.  
It is based on the physical  argument
that at high enough temperatures the modes of the system become
independent and may be represented as Gaussian fluctuations.
The argument is independent of dimension and one of the elementary
consequences of this theory is that at infinite temperature the
autocorrelation of a spin which obeys a conservation law has the long
time behavior in dimension $d$
\begin{equation}
S(t)\sim At^{-d/2}.
\label{dif}
\end{equation}
This spin diffusion phenomenology
is extensively used to study neutron scattering in real magnetic
systems~\cite{rsvw}.

It is of considerable interest to
demonstrate that this phenomenology follows in some degree 
of universality for a large class of conservative systems 
specified by a Hamiltonian and over the years there have been a
variety of arguments made for its validity~\cite{rmk,rkm,rbm,rfor}. 
In parallel to these investigations there has been a
long series of  attempts to find a
system sufficiently simple to actually compute an autocorrelation
function and to then see whether or not the spin diffusion behavior
actually holds. Most of the studies have been in one dimension
where the long time dependence of ~(\ref{dif})~ is
\begin{equation}
S(t)\sim At^{-1/2}
\label{difone}
\end{equation}

The most studied quantum mechanical model is
the spin 1/2 $XXZ$ chain of $N$ sites with periodic boundary conditions
specified by the Hamiltonian
\begin{equation}
H={1\over 2}\sum_{i=1}^{N}\left(\sigma^x_i \sigma^x_{i+1}+
\sigma^y_i \sigma^y_{i+1}+{\Delta}\sigma^z_i \sigma^z_{i+1}\right)
\label{ham}
\end{equation}
where $\sigma^j_i$ is the $j=x,y,z$ Pauli spin matrix at site $i.$
The autocorrelation functions at infinite temperature are defined as 
\begin{equation}
S^j(t;\Delta)=<\sigma^j_0(t)\sigma^j_0(0)>=\lim_{N\rightarrow \infty}
2^{-N}{\rm Tr} 
e^{-itH}\sigma^j_0 e^{itH}\sigma^j_0.
\label{defn}
\end{equation}
These correlations have been studied  
for the isotropic case $\Delta=1$ by numerical and approximation methods 
for over 30 years
~\cite{rcra,rcrb,rmb,rsl,rrmp,rbs,rbl,rbvsm,rblhvsm,rvm,rfls}.

What is very surprising is that after so much effort the answer to the
question of whether or not there is spin diffusion in the $XXZ$ model
has not been resolved (see chapter 10 of ~\cite{rvm}~for a detailed
discussion of the situation as of 1994). This is all the more
surprising since the spin 1/2 $XXZ$ model is the oldest known solvable
spin chain.
 However, after decades of effort the only exact results
known for the autocorrelation at $T=\infty$ ~(\ref{defn})~ 
are as follows:

1)
The general result proven in~\cite{rfg}~(with a mild assumption) 
that the Fourier transform
${\tilde S}^z(\omega, \Delta)$ of
$S^z(t,\Delta)$ must diverge when $\omega \rightarrow 0$ 
at least as rapidly as $\ln \omega^{-1}$ 
and that if the asymptotic behavior ~(\ref{difone})~holds then as $\omega\sim
0$
\begin{equation}
{\tilde S}^z(\omega)\sim A'\omega^{-1/2};
\label{ftom}
\end{equation}
and 

2) The results specific to $\Delta=0$ that~\cite{rnm} 
\begin{equation}
S^z(t;0)=[J_0(2t)]^2=\left[ {1\over  \pi}\int_{0}^{\pi}\cos 2t \sin
\theta d\theta\right]^2
\label{nem}
\end{equation}
where $J_0(2t)$ is the Bessel function of order zero
and ~\cite{rsjl,rbj,rcp}
\begin{equation}
S^x(t,0)=e^{-t^2}.
\label{perk}
\end{equation}

The total spin component $\sum_{i=0}^{N}\sigma^x$ will only commute 
with $H$ if $\Delta=1$ and hence in general depends on time. 
Thus $S^x(t,\Delta)$ is not expected to have spin diffusion behavior
for $\Delta\neq 1. $ However $\sum_{i=1}^{N}\sigma^z_i$ does commute with $H$
and hence does satisfies the conservation law  needed for spin diffusion. 
Nevertheless
for large $t$ we see from ~(\ref{nem})~that $S^z(t;0)$ behaves as
\begin{equation}
S^z(t;0)\sim {1\over \pi t}\cos^2(2t-{\pi\over 4})={1\over
2\pi t}\left[1+\cos(4t-{\pi\over 2})\right]
\label{exass}
\end{equation}
which is certainly not of the form~(\ref{difone}).

In the absence of exact computations recourse has been made to a
variety of speculations. For example~\cite{rmc,rczp}~it may 
be argued that the 
reason the XXZ model is solvable in the first place is because it
possesses an infinite number of conservation laws and that this violates
the assumption of independence of modes used in ~\cite{rdg,rdgb}. This
would argue that there is no reason that there should ever be spin
diffusion in the model.

One can also attempt to gain inspiration from low temperature and
field theory limit computations. As far back as ~\cite{rmb}~ and most
recently in ~\cite{rnar}, it is argued that at low temperature 
there is no spin diffusion for
$-1\leq \Delta\leq 1,$ where there is no long range order at $T=0$. 
However from a low temperature computation of Sachdev and
Damle ~\cite{rsac,rsacb}~it can
be  argued that for $1<\Delta,$ where there is long range order at
$T=0,$ spin diffusion will hold.
Thus, as a possibility, it can be suggested that for the spin$1/2$
$XXZ$ model there could be a nonanalytic behavior at $\Delta=1$ such
that at $T=\infty$ there would be spin diffusion for $\Delta>1$
and no
spin diffusion for $0\leq \Delta \leq 1.$

In this paper we assess the merits of these suggestions by an
extension of the work of ~\cite{rfls}. In particular for a spin chain of
size $N=16$ we compute exactly by diagonalizing the finite size
matrices the spin correlation functions ~(\ref{defn}). These results are given
in sec. 2. We then analyze these
results in sec. 3 in terms of the Ansatz on the long time behavior of
~\cite{rrmp}~and conclude with a discussion in sec. 4

The conclusion of this analysis as given in tables 1 and 2 
is that we do indeed see evidence 
for $0\leq \Delta \leq 1$ that
spin diffusion does not hold. Indeed it is consistent with our
data that as $t\rightarrow \infty$
\begin{equation}
S^z(t,\Delta)\sim A/t~~~{\rm for}~0\leq \Delta<1
\label{conj}
\end{equation}
which is the same as the ballistic behavior of the $\Delta=0$ (free
fermion case).
Moreover at $\Delta=1$ we find strong evidence supporting the behavior
\begin{equation}
S^z(t,1)\sim A/t^{.705}
\label{iso}
\end{equation}
which is neither ballistic nor diffusive. This result  is in agreement
with the estimate of ~\cite{rvm}~that ${\tilde S}^z(\omega)$ diverges as
$\omega^{-.37\pm .12}$ as $\omega \rightarrow 0.$ 
For $\Delta>1$ we cannot be so positive as to whether or not spin
diffusion exists  and we defer further
discussion to the end of sec. 3. 
\section{Finite Size results for $N=16$}

We have evaluated $S^z(t,\Delta)$ for chains up to size $N=16$ using
the methods of exact diagonalization presented in \cite{rfls}.
We plot our results for $0 \leq \Delta \leq 1.3$ in steps of $0.1$
 in Fig. 1.
 The results
presented here extend the previous work principally in that data is
used which range over a larger time interval and many more values of
$\Delta$ have been studied.

Data on finite size systems will only well approximate infinite size
systems for some finite time interval. In \cite{rfls}~the criteria used was
that the data for $N=14$ and $N=16$ should closely agree. This
amounts to using the $N=16$ data to tell us how much of
the $N=14$ data can be used. However the $N=16$ data will agree with
the $N=\infty$ curve for a longer time interval. We estimate this
larger interval in two ways. First by comparing the exact result~(\ref{nem})~ 
for $\Delta=0$ with the finite size result for $N=16.$ The second is by
making an extrapolation using the data for $N=12,~14,$ and $16.$ We
make these comparisons in Fig. 2 for $\Delta=0.$  This
allows us to effectively extend the range of $t$ from $t_{ \rm
max}=5.0$ which was what was used in \cite{rfls}~to  $t_{\rm max}=6$
or greater
in some cases. The maximum time is estimated for each
$\Delta$ separately since with the normalization of (\ref{ham})~
the maximum time will
decrease as $\Delta$ increases. It is possible to introduce a
normalization which makes $t_{\rm max}$ relatively independent of
$\Delta$ but since this is somewhat {\it ad hoc} and arbitrary we will not
do this here.

We have also made a similar study for the correlation $S^x(t;\Delta).$
These results are presented graphically in Fig. 3.

There are two qualitative features of these graphs which should be
noted. First of all in the graph for $S^x(t,\Delta)$ the long time
behavior of $\Delta=1$ lies higher that all other values of
$\Delta$. This is expected from fact that if $\Delta=1$ then $S^x(t)=S^z(t)$
and thus the long time behavior must be algebraic which is to be
contrasted with the exponential decay expected for all other values
of $\Delta$ where there is no conservation of $\sum_j \sigma^x_j.$
However near $t=6.0$ the curves for $\Delta=0.9$ and $\Delta=1.0$ are
so very close that this change in asymptotic behavior cannot yet be
seen in the data.

The other point to note is that in the Fig1b we see clearly that near
$t=5.0$ there is crossing of curves for $0\leq \Delta \leq 0.5.$ This
effect is real and is not an artifact of the size $N=16.$ We interpret
this as evidence that if $0\leq \Delta \leq 0.5$ then $t$ must be
greater than $5.0$ before the true long time asymptotic regime is
seen. In  other words we take this as evidence that there is a
crossover in the system.

\section{Asymptotic fitting} 

In order to analyze the existence of spin diffusion we must extract
the long time behavior of the correlation function ~$S^z(t,\Delta).$
It is obvious from  Fig.1 that for times up to $t=6$ that there
are oscillations in the data and that a simple power law will not
be adequate to describe the results. Here we confront this problem by
fitting the curves with an extension of the simple Ansatz 
proposed in ~\cite{rrmp}~which
incorporates a decaying oscillation as well as a decay with an
arbitrary power law
\begin{equation}
f(t;d,A,B,\gamma,\Omega,t_0)=t^{-d}[A+Be^{-\gamma (t-t_0)}\cos\Omega
(t-t_0)].
\label{fit}
\end{equation}
In Figs. 4,5 and 6 we illustrate this 
fitting procedure by showing a least squares fit of 
\begin{equation} g_f(t;B,\gamma,\Omega,t_0)=Be^{-\gamma (t-t_0)}\cos\Omega
(t-t_0)
\label{gfit}
\end{equation}
to the function obtained from the $N=16$ data of fig.1
\begin{equation}g_s(t,d,A;\Delta)=t^d S^z(t;\Delta)-A
\label{gdata}
\end{equation}
for the values $\Delta=.5,~1.0$ and $1.3.$

More systematically we present in Table 1 the parameters of the form
~(\ref{fit})~ which best fit all the data with $N=16$ presented in Fig. 1 
and the time intervals over which the fit is valid. In Table 2 we give the
corresponding fitting parameters and time intervals for the data with
$N=14.$ We note from ~(\ref{exass}) that at $\Delta=0$ the exact values
of the parameters are
\begin{equation}
d=1~~A=B={1/2\pi}=.159155\cdots,~~\gamma=0,~~\Omega=4.0,~~t_0=5\pi/8=1.9635\cdots
\label{exact}
\end{equation}

It is very instructive to compare the values of the 
fitting parameters
as obtained in the two tables. For  $\Delta=0$ and $1$ the values of
$d$ changes little in going from $N=14$ to $N=16.$ But this is not the
case for all values of $\Delta$ and we compare the cases $N=14$ and
$N=16$ in Fig. 7. First of all we see that for $0.4 \leq \Delta \leq
0.9$ there is a great deal of variation in $d$ as we go from
$N=14$ to $N=16.$ We interpret this as an indication that there is a
crossover in the behavior from small to large times. We  expect that
as $N$ increases this trend will continue for all $0< \Delta < 1.0$
and  eventually the data for $0< \Delta <.4$ will be affected also.
We cannot definitively say from these plots what the true value of $d$
will be for $N\rightarrow \infty$ but because the fitted value of $d$
is increasing in our tables for $0 < \Delta<1$ we conclude that the spin
diffusion value of $1/2$ is never attained. At $\Delta=1$ the value
 $d=.705$ is seen to be quite stable and we note that to within our
accuracy it is equal to $2^{-1/2}.$ From our analysis $d$
could either be a continuous function of $\Delta$ at $\Delta=1$ or we
could have $d=1$ for $0\leq \Delta <1.$

For $1<\Delta$ we cannot be so positive in our conclusion. For
$\Delta=1.2,~1.3$ the fitted values of $d$ do indeed decrease as $N$
goes from 14 to 16 and it is not out of the question that as
$N\rightarrow \infty$ the limiting value could be 1/2 for all values
of $1<\Delta.$  However this is not mandated by our results.

\section{Discussion}
 
The conclusion of the previous section is that in the spin 1/2 XXZ
model specified by ~(\ref{ham})~there is no spin diffusion for $0\leq \Delta
\leq 1.$ This needs to be discussed both as
to its correctness and its implications.

We first acknowledge that it is possible to make contradictory
suggestions as to what is to be expected and that different authors
seem to implicitly start from different conceptions of what is going
on. For example the authors in the mid sixties speak of their work as
the hydrodynamic approximation and thus is sometimes said to be
inapplicable to one and two dimensions where the Fourier transform of
the autocorrelation function diverges. This would argue that one
dimension always needs a separate treatment. However, most of the more
recent authors seem to write as if the diffusion form (\ref{difone})~is to be
expected in one dimension if only in the fact that deviations from it
are called ``anomalous.''

Indeed, the question of whether or not this anomalous diffusion occurs in
the classical one dimensional Heisenberg model has been a subject of
some controversy in the past 10
years~\cite{rgma,rgl,rgmb,rglb,rdr,rbgl}. It is agreed by all
that for times up to about 50 an exponent of $d \sim .61$ can be
obtained from computer simulations. What is controversial is the
ultimate long time behavior. The arguments are summarized in
~\cite{rvm}. What is important for us here is to note that for the spin 1/2
quantum mechanical case it is not possible with current computing
power to go to anything approaching the large time of these classical
computations. Thus all  speculations and conjectures about the long
time behavior of the spin 1/2 model are subject to the proviso that we
assume that times up to $t=6$ are able to teach us about the true
$t\rightarrow \infty$ behavior. It is our belief that the
integrability of the spin 1/2 XXZ chain will rule out the possibility
of a new time scale appearing for large times but there is no way of
verifying this short of an exact computation.

If it is accepted that the estimate of the asymptotic behavior we have
presented here is indeed correct for the model (\ref{ham})~it must then be
asked whether or not this is generic in any sense. It is here that
the question of the relation of an integrable to a generic system
needs to be addressed. 

In the first place there is ample computer evidence~\cite{rpzbmm}~that if a
sufficiently strong
next nearest neighbor interaction is added to ~(\ref{ham})~then the level
spacing statistics will change from Poisson for ~(\ref{ham})~to those of the
GOE ensemble of random matrices. It thus might be supposed that this
will change the long time behavior of the correlation functions. We
have indeed looked at this in the $N=16$ system but find that the
asymptotic behavior up to time $t=6$ does not change. But, of course,
this proves little or nothing since if a scale is opened up at larger
times the $N=16$ system can hardly be expected to see it. It is
certainly possible that all the complications seen in the classical
system can occur for the non integrable quantum spin chain if we could
go to large enough times. This is the place where ideas of quantum
chaos should be able to intersect many body condensed matter physics.

We also comment further on our suggestion of nonanalytic
behavior at $\Delta=1.$ It is of course perfectly reasonable that at
T=0 there will be a marked difference in the physics of $0\leq \Delta
\leq 1$ and $1<\Delta.$ In the first case there is no long range order
and no gap in the spectrum while in the second case there is both long
range order and a gap in the spectrum~\cite{ryy,rjkm}. 
It is exactly these qualitative
differences which feature in the low temperature  computations of
~\cite{rnar} and ~\cite{rsac,rsacb}. But at high temperature 
it seems unreasonable that
such low temperature properties as whether or not there is a gap in
the excitations above the ground state should make any difference. It
would seem that to maintain that there is nonanalytic 
behavior at ~$\Delta=1$ we must violate our physical intuition.  

The resolution to this would seem to lie in the integrability of
(\ref{ham}). Indeed the thermodynamics have been studied by means of the
thermodynamic Bethe's Ansatz method and a set of two coupled
nonlinear integral equations has been derived
~\cite{rklu,rddv}~ whose solution gives
the free energy at all $T.$ These integral equations have the feature that
they do take two distinct forms depending on which of the two regimes
$\Delta$ lies in. In a more picturesque fashion we can say that the
integrability of the model extends the low temperature description
in terms of particles to all temperatures. For this reason we expect that
the dynamics at infinite temperature of the spin 1/2 XXZ model are not
generic. The study of the crossover from integrable to generic
behavior as next nearest neighbor interactions are added would
constitute a major step towards formulating what should be called a
quantum KAM theorem. Such a theorem would go a long way towards
clarifying the status of diffusion at high temperatures in quantum
mechanical systems.

Finally we remind the reader that because of the integrability of the
spin 1/2 XXZ chain it is firmly to be expected that the time dependent
correlations studied in this paper can be exactly evaluated. We
hope that the numerical results presented here will stimulate the
analytic solution of this problem. 
\acknowledgments

We are pleased to acknowledge useful discussions with  A. Kl{\"u}mper
and S. Sachdev.
This work is supported in part by the National Science Foundation
under grant DMR 97-03543.

\begin{table}[b]
\begin{tabular}{|c|c|c|c|c|c|c|c|c|c|}
&&&&&&&&&\\
$\Delta$ &$d$ & $A$ & $B$ & $\gamma$ & $\Omega$& $t0$ & $t1$ & $t2$ & $\chi^2$\\
&&&&&&&&&\\
0.0 &  1.000 &  0.159 &  0.159 &  0.000 &  4.010 &  1.976 &  2.5 &  5.0 &   1.199e-07\\
0.1 &  0.961 &  0.156 &  0.153 &  0.023 &  4.009 &  1.977 &  2.5 &  5.0 &   1.651e-07\\
0.2 &  0.875 &  0.152 &  0.141 &  0.083 &  4.007 &  1.976 &  2.5 &  5.0 &   1.520e-07\\
0.3 &  0.810 &  0.156 &  0.131 &  0.159 &  4.007 &  1.976 &  2.5 &  5.0 &   1.071e-08\\
0.4 &  0.835 &  0.180 &  0.127 &  0.217 &  4.039 &  1.985 &  3.0 &  5.8 &   1.016e-06\\
0.5 &  0.912 &  0.219 &  0.120 &  0.246 &  4.080 &  1.987 &  3.5 &  6.0 &   7.327e-08\\
0.6 &  0.941 &  0.246 &  0.108 &  0.290 &  4.138 &  1.987 &  3.5 &  6.0 &   1.659e-07\\
0.7 &  0.902 &  0.251 &  0.092 &  0.361 &  4.213 &  1.989 &  3.1 &  6.0 &   4.273e-06\\
0.8 &  0.840 &  0.249 &  0.079 &  0.453 &  4.274 &  1.973 &  3.1 &  6.0 &   5.789e-07\\
0.9 &  0.771 &  0.248 &  0.066 &  0.558 &  4.353 &  1.958 &  3.0 &  5.7 &   2.165e-07\\
1.0 &  0.705 &  0.247 &  0.053 &  0.668 &  4.439 &  1.933 &  3.2 &  5.1 &   4.202e-10\\
1.1 &  0.646 &  0.249 &  0.041 &  0.775 &  4.568 &  1.906 &  3.0 &  4.5 &   9.062e-11\\
1.2 &  0.602 &  0.254 &  0.027 &  0.811 &  4.779 &  1.866 &  3.0 &  4.5 &   8.756e-11\\
1.3 &  0.572 &  0.263 &  0.015 &  0.669 &  5.010 &  1.769 &  3.0 &  4.5 &   8.568e-11\\
\end{tabular}
\caption{The best fit parameters for the system with $N=16$ spins. The
entries $t_1$ and $t_2$ indicate the time interval for which the fit
(\ref{fit}) is good. }
\label{one}
\end{table}
\begin{table}[b]
\begin{tabular}{|c|c|c|c|c|c|c|c|c|c|}
&&&&&&&&&\\
$\Delta$ &$d$ & $A$ & $B$ & $\gamma$ & $\Omega$& $t0$ & $t1$ & $t2$ & $\chi^2$\\
&&&&&&&&&\\
0.0 &  0.985 &  0.156 &  0.157 &  0.006 &  4.015 &  1.978 &  2.0 &  4.0 &   4.283e-08\\
0.1 &  0.954 &  0.155 &  0.153 &  0.028 &  4.013 &  1.978 &  2.0 &  4.0 &   2.255e-08\\
0.2 &  0.882 &  0.154 &  0.142 &  0.085 &  4.008 &  1.977 &  2.0 &  4.0 &   1.474e-09\\
0.3 &  0.809 &  0.156 &  0.131 &  0.158 &  4.006 &  1.975 &  2.0 &  4.0 &   1.552e-09\\
0.4 &  0.784 &  0.168 &  0.123 &  0.231 &  4.013 &  1.973 &  2.5 &  4.8 &   1.653e-07\\
0.5 &  0.833 &  0.196 &  0.123 &  0.301 &  4.055 &  1.979 &  2.8 &  5.0 &   5.627e-07\\
0.6 &  0.885 &  0.227 &  0.114 &  0.348 &  4.134 &  1.990 &  3.0 &  5.2 &   4.273e-07\\
0.7 &  0.840 &  0.230 &  0.103 &  0.446 &  4.145 &  1.960 &  2.5 &  5.1 &   9.297e-06\\
0.8 &  0.801 &  0.236 &  0.086 &  0.515 &  4.210 &  1.948 &  2.5 &  5.0 &   2.759e-06\\
0.9 &  0.746 &  0.239 &  0.071 &  0.599 &  4.285 &  1.932 &  2.5 &  4.8 &   5.122e-07\\
1.0 &  0.694 &  0.244 &  0.056 &  0.692 &  4.403 &  1.922 &  2.8 &  4.5 &   4.914e-09\\
1.1 &  0.640 &  0.247 &  0.041 &  0.773 &  4.536 &  1.895 &  2.6 &  4.0 &   5.642e-11\\
1.2 &  0.608 &  0.256 &  0.027 &  0.821 &  4.804 &  1.875 &  2.6 &  4.0 &   2.236e-10\\
1.3 &  0.577 &  0.265 &  0.015 &  0.654 &  5.025 &  1.774 &  3.0 &  4.0 &   4.089e-12\\
\end{tabular}
\caption{The best fit parameters for the system with $N=14$ spins. The
entries $t_1$ and $t_2$ indicate the time interval for which the fit
(\ref{fit}) is good.}
\label{two}
\end{table}

\newpage


Fig. 1a. The correlation function $S^z(t,\Delta)$ as computed from the
$N=16$ spin chain for $0\leq \Delta \leq 1.3$ in steps of $.1$ On this
scale the curves are monotonic functions of $\Delta$ with $\Delta=0$
lying lowest. The curve for $\Delta=1$ is dashed.

Fig. 1b. The correlation function $S^z(t,\Delta)$ of Fig. 1a on the time range 
$1\leq t \leq 6$. The curve for $\Delta=1$ is dashed and for
$\Delta=0.5$ is dot-dashed.
 On this scale we see that near $t=5.0$ that for
$0\leq \Delta \leq .5$ that there is crossing of the curves.

Fig. 2. Comparison of the exact result~\ref{nem}~$tS^z(t,0)$ (solid line) 
with the finite
size computations for $N=12$ (long dashes), $N=14$ (dot-dashes) and
$N=16$ (short dashes). 

Fig. 3a. The correlation function $S^x(t,\Delta)$ as computed from the
$N=16$ spin chain for the values $\Delta=0.0,~0.1,~0.3,~0.5,~0.7,~0.9,~
1.0,~1.1,~1.3.$ The curve
$\Delta=1$ is dashed and for $0 \leq t \leq 1.2$ the curves are
monotonic in $\Delta$ with $\Delta=0$ lying highest.

Fig. 3b.The correlation function $S^x(t,\Delta)$ of Fig. 3a on the time
interval $1\leq t \leq 6.$ The values of $\Delta$ are indicated on the
curves.

Fig. 4. Least squares fit (solid line) of the $N=16$ data (vertical crosses) 
for $S^z(t,\Delta)$ in the form ~(\ref{gdata})  
to the form ~(\ref{gfit})~ for $\Delta=.5.$ 
The values of the parameters are
$d=0.912,A=0.219,B=0.120,\gamma=0.246,\Omega=4.080, t_0=1.987,$ 
and the interval of best fit is
$3.5<t<6.0.$ The data for $N=14$ (diagonal crosses) and $N=12$ (triangles) 
is also shown.

Fig. 5. Least squares fit (solid line) of the $N=16$ data (vertical
crosses) for $S^z(t,\Delta)$ in the form ~(\ref{gdata})  
to the form ~(\ref{gfit})~ for $\Delta=1.0.$
The values of the parameters in are
$d=0.705,A=0.247,B=0.053,\gamma=0.668,\Omega=4.439, t_0=1.933,$ 
and the interval of best fit is
$3.2<t<5.1.$ The data for $N=14$ (diagonal crosses) and $N=12$ (triangles) 
is also shown.

Fig. 6. Least squares fit (solid line) of the $N=16$ data 
(vertical crosses) for $S^z(t,\Delta)$ in the form ~(\ref{gdata})  
to the form ~(\ref{gfit})~  
for $\Delta=1.3.$ 
The values of the parameters are
$d=0.572,A=0.263,B=0.015,\gamma=0.669,\Omega=5.010, t_0=1.769,$ 
and the interval of best fit is
$3.0<t<4.5.$ The data for $N=14$ (diagonal crosses) and $N=12$ (triangles) 
is also shown.

Fig. 7. Comparison of the best fit values of $d$ for $N=14$ and $N=16.$
\newpage

\centerline{\epsfxsize=6in\epsfbox{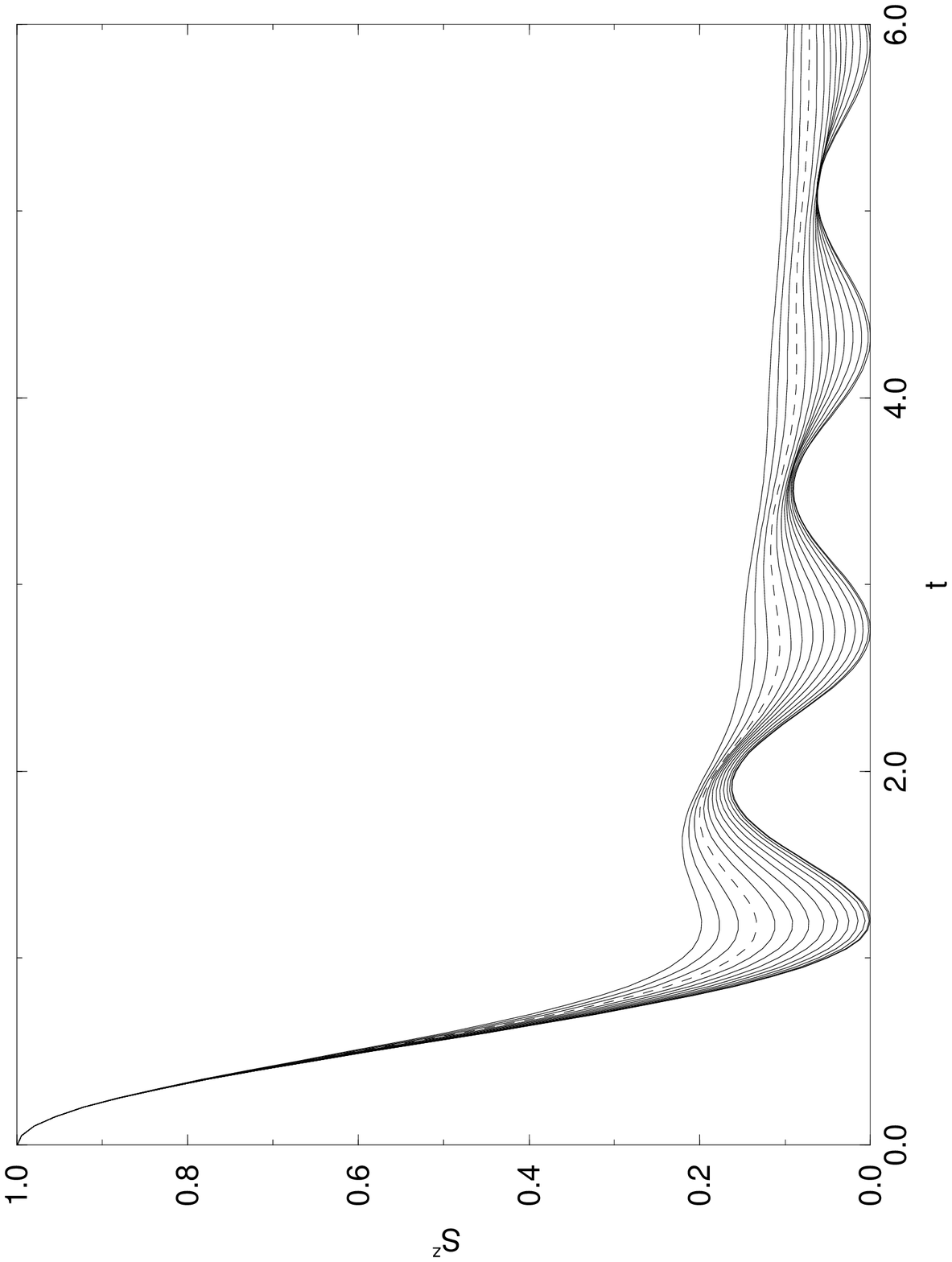}}
\centerline{Fig.1a}

\centerline{\epsfxsize=6in\epsfbox{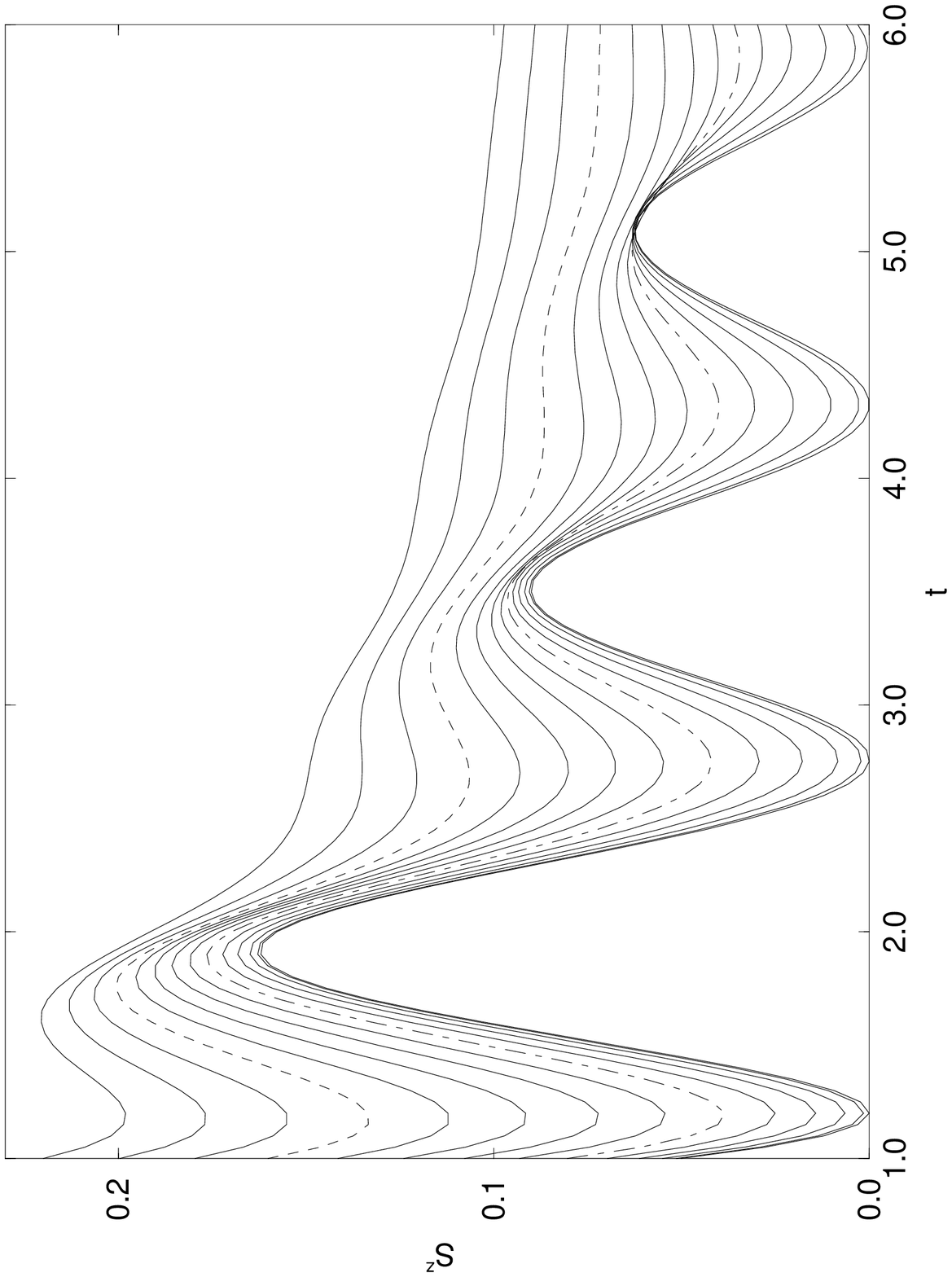}}
\centerline{Fig.1b}

\centerline{\epsfxsize=6in\epsfbox{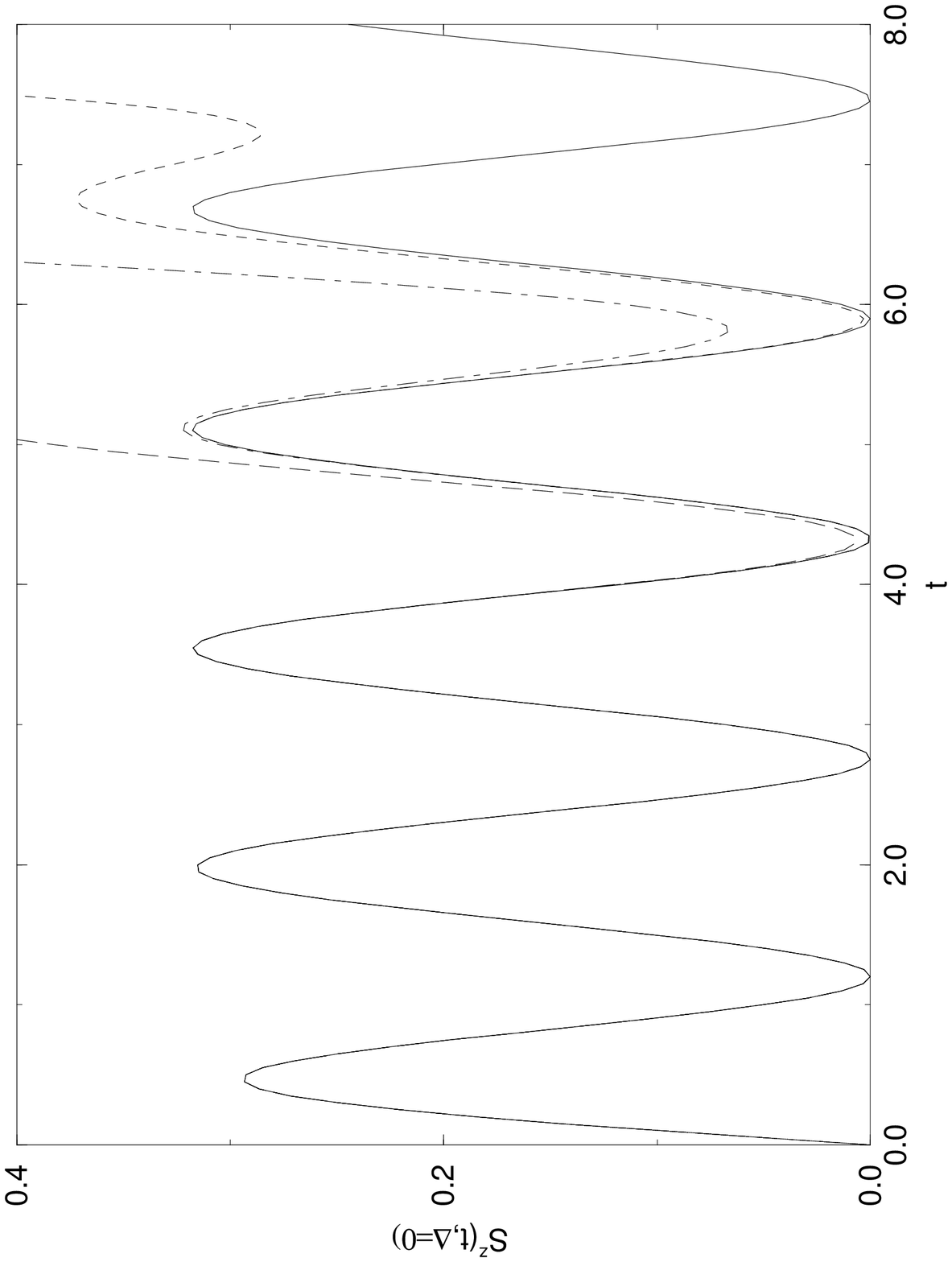}}
\centerline{Fig.2}

\centerline{\epsfxsize=6in\epsfbox{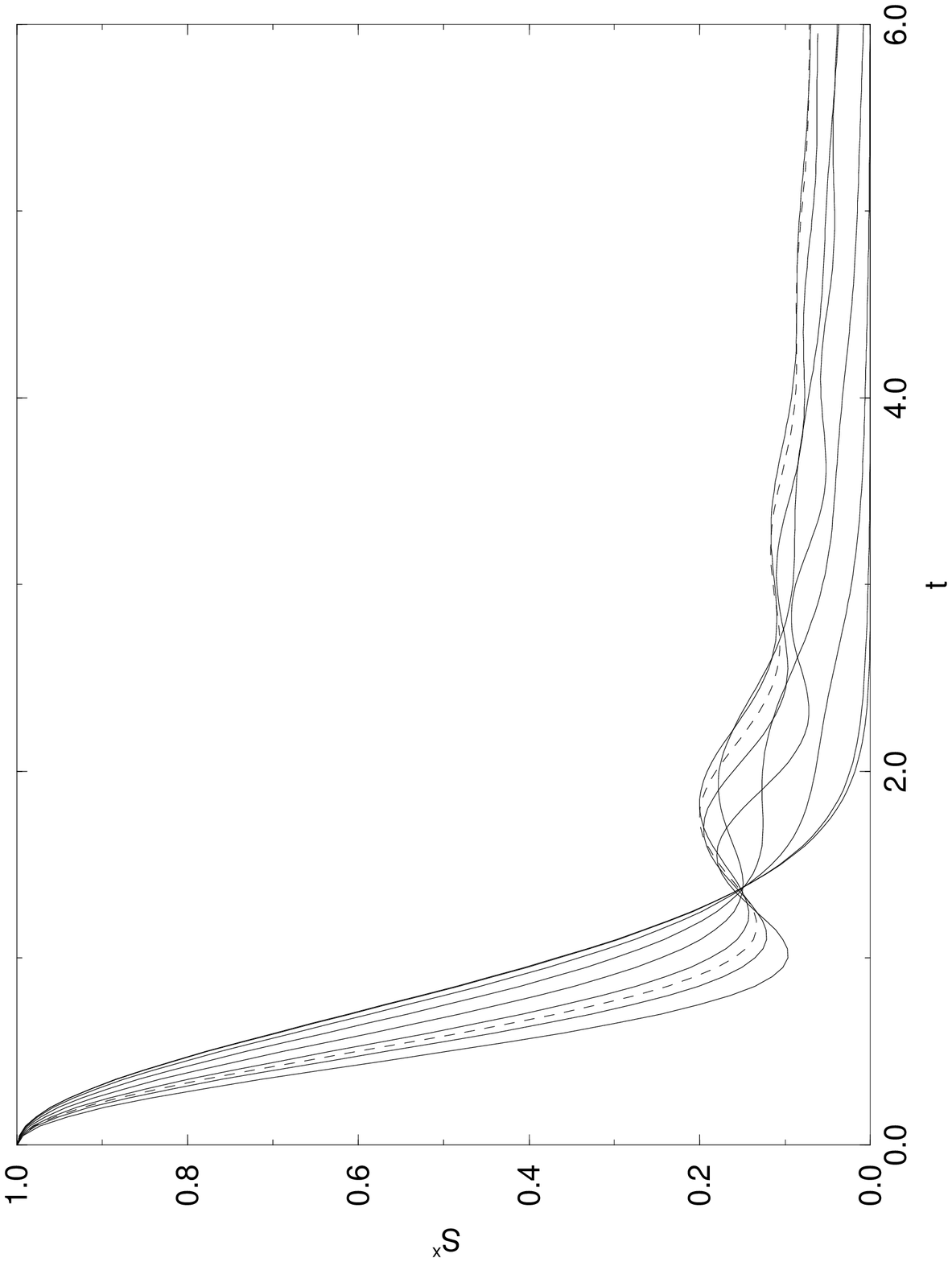}}
\centerline{Fig.3a}
\centerline{\epsfxsize=6in\epsfbox{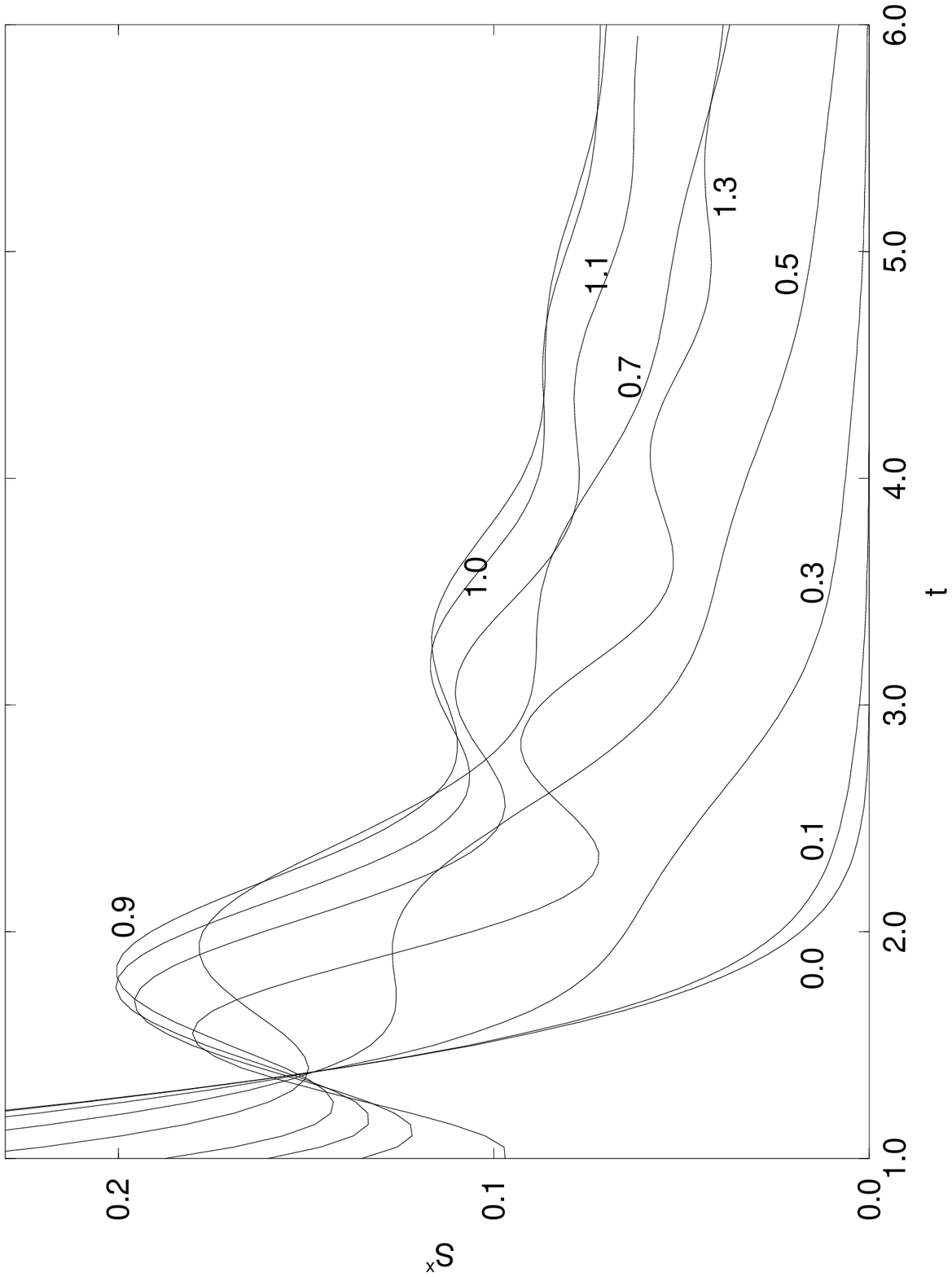}}
\centerline{Fig.3b}

\centerline{\epsfxsize=6in\epsfbox{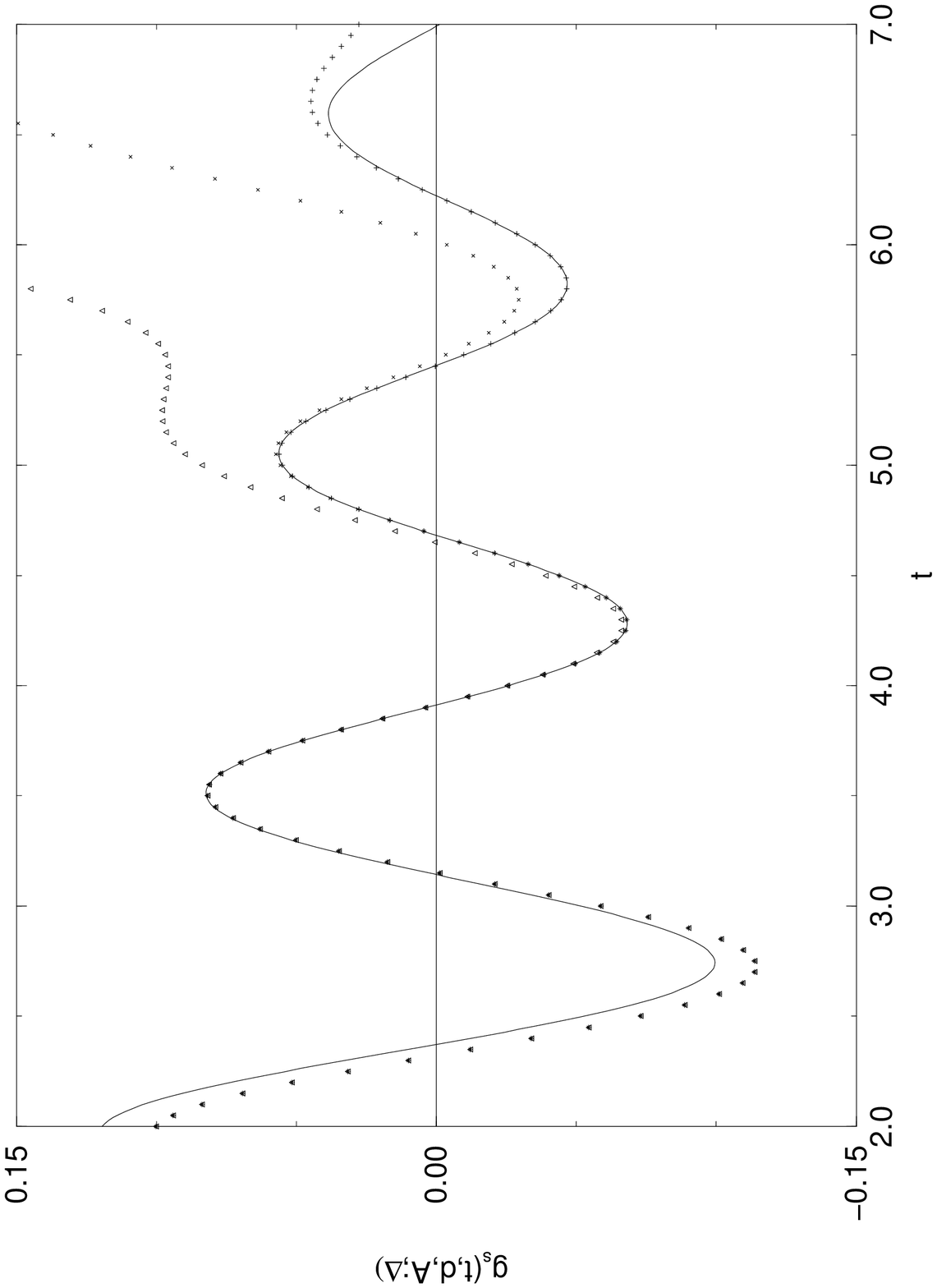}}
\centerline{Fig.4}

\centerline{\epsfxsize=6in\epsfbox{ 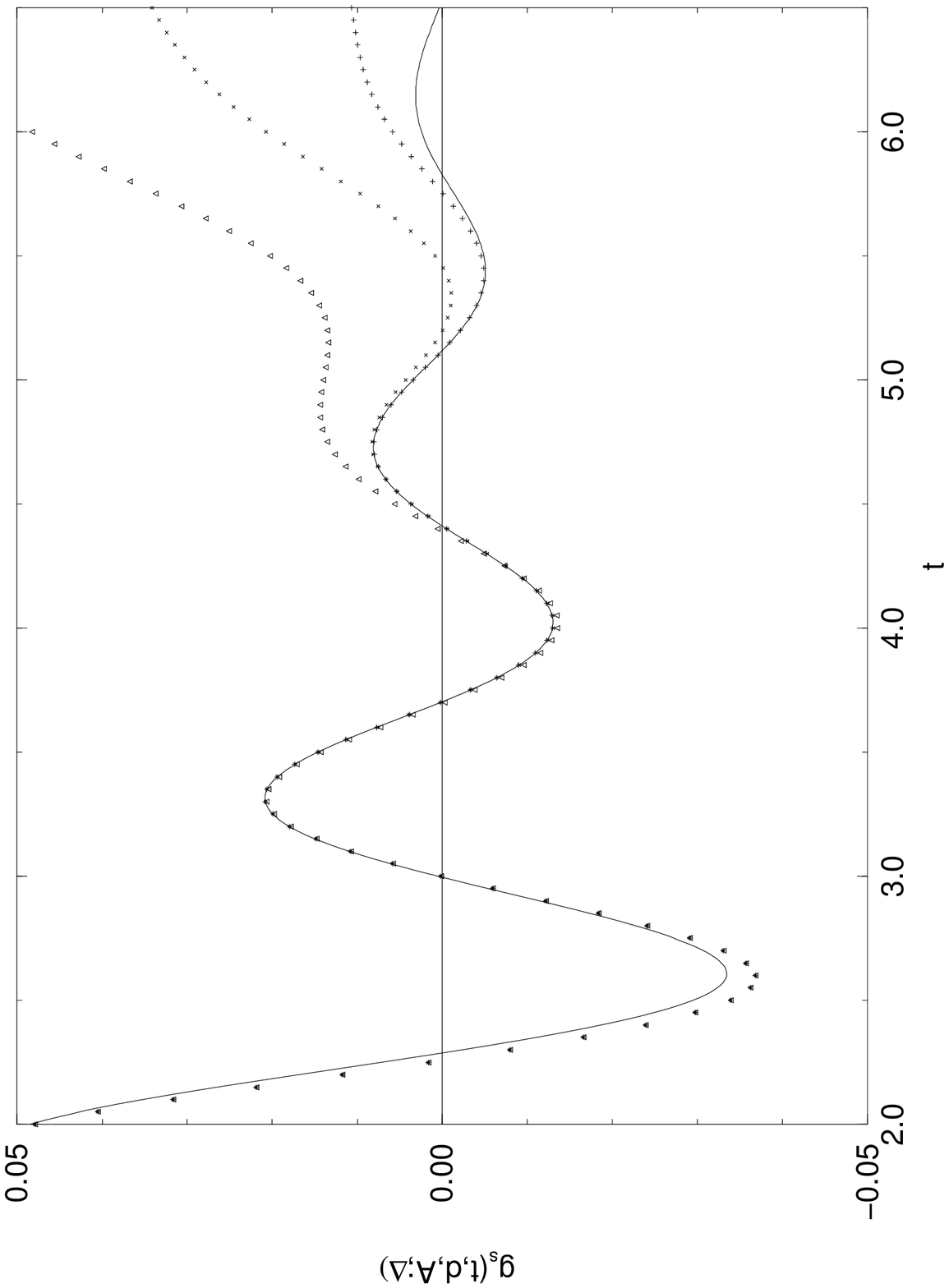}}
\centerline{Fig.5}

\centerline{\epsfxsize=6in\epsfbox{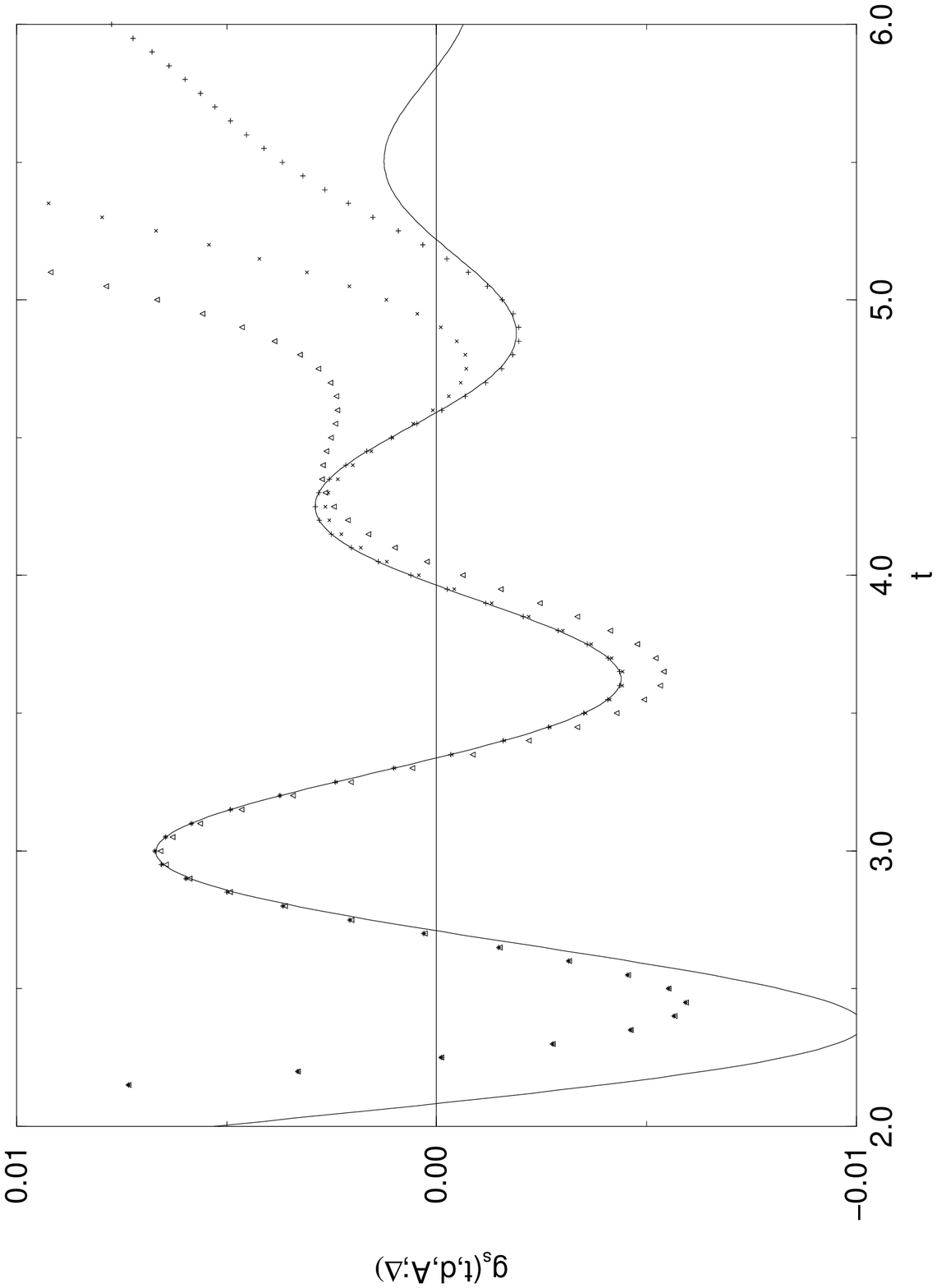}}
\centerline{Fig.6}

\centerline{\epsfxsize=6in\epsfbox{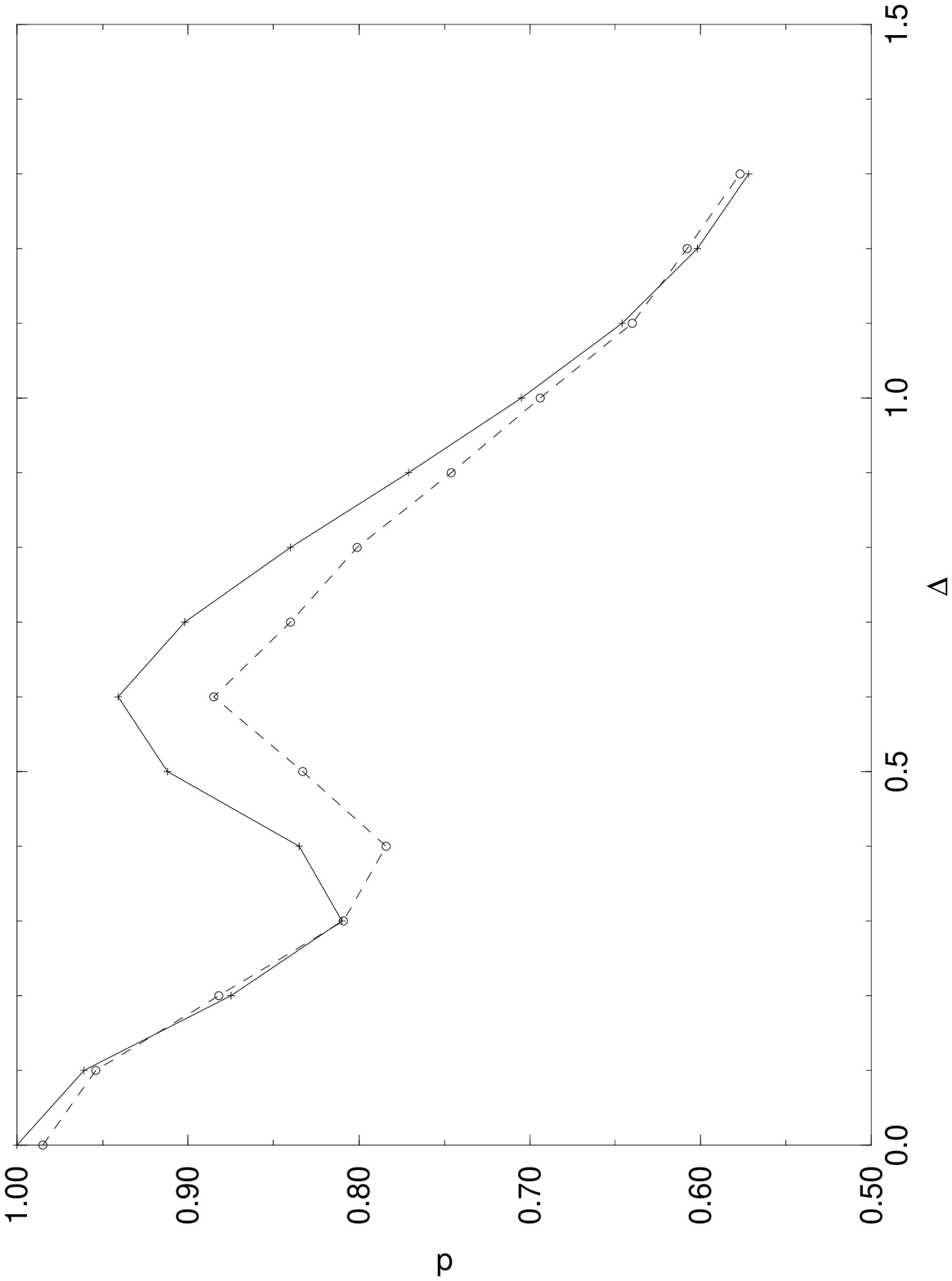}}
\centerline{Fig.7}
\end{document}